\documentclass[%
 amsmath,amssymb,
 aps,
]{revtex4-2}

\usepackage{graphicx}
\usepackage{bm}
\usepackage{float}

\begin{document}

\title{Estimation of posture and joint angle of human body using foot pressure distribution :Morphological computation with human foot}

\author{Yo Kobayashi}
\email{yo.kobayashi@es.osaka-u.ac.jp}%
\affiliation{Graduate School of Engineering Science, Osaka University, Osaka, Japan}%

\author{Yasutaka Nakashima}
\affiliation{Faculty of Engineering, Kyushu University, Fukuoka, Japan}%


\begin{abstract}

This paper proposes a novel contact and wearable sensing system for estimating the upper body posture and joint angles (ankle, knee, and hip) of the human body using foot pressure distribution information obtained from a sensor attached to the plantar region. In the proposed estimation method, sensors are installed only on the plantar region, which is the end of the human body and the point of contact with the environment. The posture and joint angles of other parts of the body are estimated using only this information. As a contact and wearable sensor, the proposed system differs from previous measurement systems in the sense that the sensor does not need to be placed near the target joint or body. The estimation was carried out using a multivariate linear regression model with the foot pressure distribution as the input and the joint angle or posture as the output. The results reveal that it is possible to estimate the posture and joint angles of the human body from foot pressure distribution information (R2$\fallingdotseq$0.9). The proposed estimation method was validated by morphological computation to confirm that it is enabled by foot morphology. The validation approach compared the estimation accuracy achieved when an object was interposed between the foot pressure distribution sensor and the plantar region and the morphological relationship of the plantar region to the environment varied. The results reveal that there is a significant difference in the estimation accuracy between cases with and without an intervening object, suggesting that the morphology of the plantar region contributes to the estimation. Furthermore, the proposed estimation method is considered as physical reservoir computing, wherein the human foot is used as a computational resource. 


\end{abstract}

\maketitle

\section{\label{sec:introduction}Introduction}

The measurement and estimation of the joint angles and posture of the human body play an important role in various fields. For example, the estimation of the human body posture and joint angles is an important research topic in various fields such as robotics, human–computer interaction, and biomechanics \cite{woods2022joint, williamson2001detecting,el2022virtual,rajasekaran2015adaptive, 9952586,yang2019gesture,yang2019gesture, bonnet2015monitoring, rajasekaran2015adaptive,el2012shoulder, el2015human,majumder2020wearable, 9141313,teague2020wearable}.
In the field of robotics, the accurate measurement and estimation of the human body posture and joint angles is important. For example, when a human and a robot work together, the human’s posture and movement must be accurately determined \cite{el2022virtual}. Additionally, in the case of wearable assistive robots, the posture and angle of the wearer when controlling the device must be estimated \cite{rajasekaran2015adaptive}.
In the field of human–computer interaction, understanding the movement and posture of the human body is important for improving human–computer interaction. For example, technologies such as virtual reality (VR) and augmented reality (AR) detect the user’s body movement and posture to reflect information on the actual environment in digital space \cite{9952586}, and can also use gestures and body movement to enable natural and intuitive interaction with computers \cite{yang2019gesture}.
The field of biomechanics considers kinematics and other aspects of the human body, and has practical application in sports, rehabilitation, and medicine. In these fields, it is important that the posture and joint angles of the human body are accurately estimated to understand human motion \cite{bonnet2015monitoring,el2012shoulder, el2015human,majumder2020wearable, 9141313,teague2020wearable}. For example, understanding the mechanical properties and functions of the musculoskeletal system in sports, work, and daily life can help in evaluating the performance of human movement, and contributes toward the diagnosis, prevention, and treatment of various disorders and injuries related to posture, balance, and mobility \cite{teague2020wearable}.

Different sensors combined with diverse information processing methods are used for joint angle and posture estimation. Generally, these sensors are categorized into contact and wearable sensors and non-contact sensing systems. Regarding non-contact sensing systems, there are two main types. The first type is motion capture using optical markers \cite{roux2002evaluation}. Optical motion capture systems have been used to quantify joint kinematics by tracking the position of reflective surface markers during dynamic activity \cite{murata2019estimation}. However, these systems are expensive, limited to controlled laboratory environments, and prone to occlusion \cite{welch2002motion}. The second type is vision-based motion capture, which uses a small number of RGB or RGB-D cameras to perform pose estimation. Vision-based methods are small systems compared with optical motion capture systems \cite{chen2020cross, habibie2019wild, mehta2020xnect, trumble2016deep, xiang2019monocular}. However, although these systems are expected to extract distinguishable features from an image, they are sensitive to human appearance and typically underperform when untextured clothing or challenging ambient lighting are present \cite{yi2021transpose}. Furthermore, vision-based methods are susceptible to occlusion.

Joint angle and posture estimation using contact and wearable sensors is independent of the measurement environment and does not suffer from the occlusion problems of non-contact sensing systems \cite{yi2021transpose}. Additionally, complex equipment is not required and there are no restrictions on the range of use \cite{yi2021transpose}. For contact and wearable sensors, research has been conducted on encoder and goniometer measurements. These contact sensors are placed directly on the joints and can perform high-precision measurements. In many applications, however, sensor placement without disturbing the joint integrity is difficult, and can be uncomfortable and cumbersome when encoders are integrated into wearable systems \cite{woods2022joint}. For other single-joint motion, a goniometer can be used \cite{edwards2004measuring}. However, goniometers require careful alignment to the joint center and must be repositioned each time the joint of interest changes, which renders them incapable of monitoring multi-joint motion \cite{bonnet2015monitoring}.
In contrast, inertial measurement units (IMUs) are placed on links at a distance from the joints, providing design flexibility and comfort without affecting the system dynamics. The IMUs are small and portable, allowing for continuous measurements without location constraints. Theoretically, the position and orientation are calculated by the simple integration of acceleration and angular velocity \cite{woods2022joint}. However, small errors in the measurement data accumulate, resulting in position and orientation errors, which is known as drift \cite{murata2019estimation}.

This study proposes a novel contact and wearable sensing system to estimate the posture and joint angles of the human body using foot pressure distribution information obtained from a sensor attached to the plantar region. In this estimation method, sensors are installed only on the plantar region, which is the end of the human body and the point of contact with the environment, and the posture and joint angles of other parts of the body are estimated only from this information. As a contact and wearable sensor, the proposed system differs from previous measurement systems in the sense that the sensor does not need to be placed near the target joint or body. In this study, estimation was carried out using a multivariate linear regression model. Additionally, the proposed estimation method was validated by morphological computation to confirm that it is enabled by foot morphology. In the validation process, the estimation accuracy achieved when an object is interposed between the foot pressure distribution sensor and the plantar region was compared with that achieved when the morphological relationship of the plantar region to the environment varied. The findings of this study demonstrate that, by processing a multivariate linear regression model utilizing foot morphology, it is possible to estimate the posture and joint angles of the human body from foot pressure distribution information.

\section{\label{methods} Methods}

\begin{figure*}
\includegraphics[width=14cm]{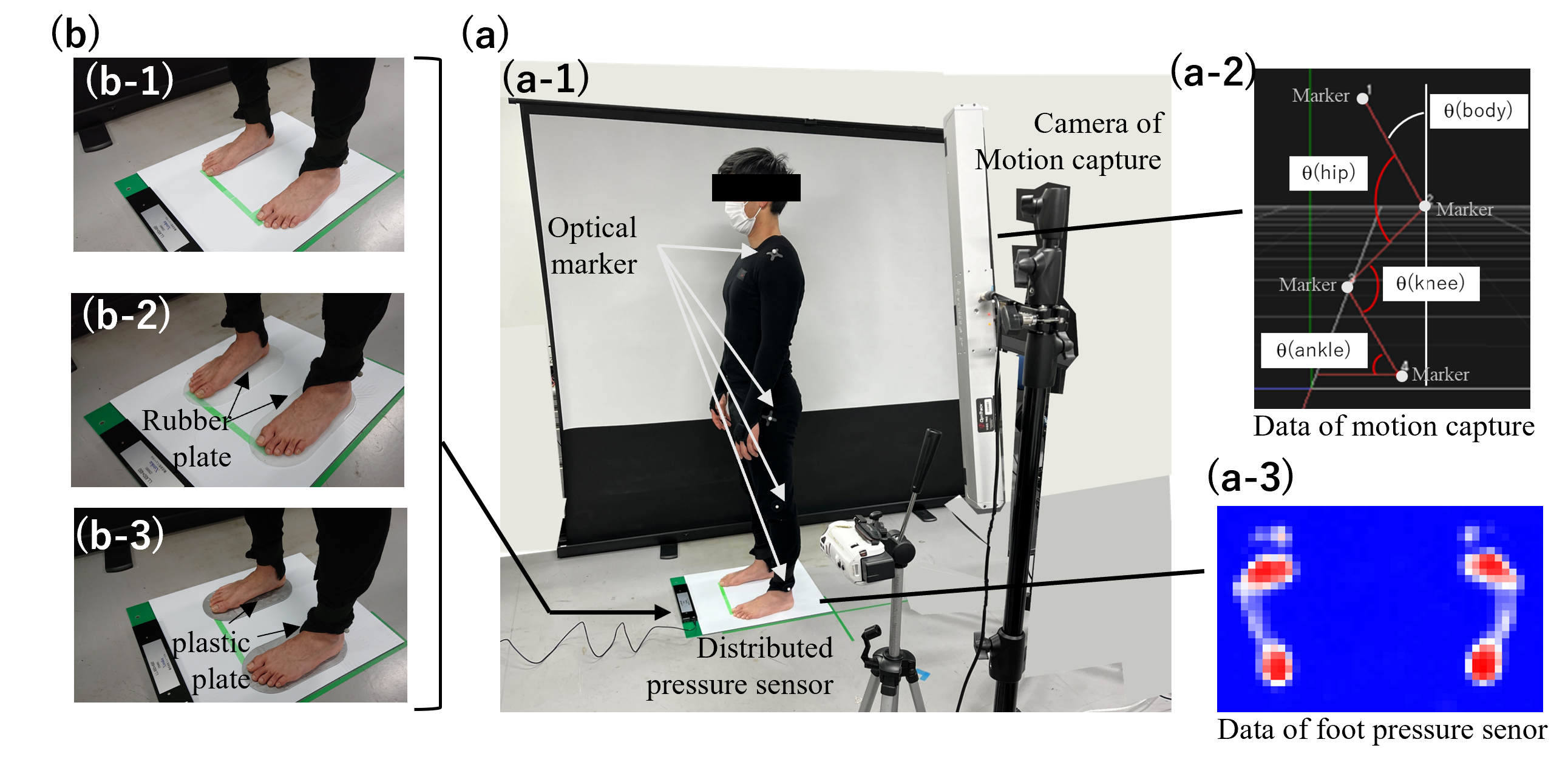}
\caption{\label{fig:concept}
Experimental setup: \textbf{(a)} (a-1) experimental setup in this study; (a-2) motion capture was used to obtain the posture and joint angles of the human body; (a-3) a sheet-shaped distributed pressure sensor was used to measure the pressure distribution on the plantar surface of the foot.  \textbf{(b)} The experiment was conducted under three conditions: (b-1) condition wherein nothing was placed between the sole and the pressure distribution sensor (Condition (A): nothing); (b-2) condition wherein a rubber plate was placed between the sole and the pressure distribution sensor (Condition (B): rubber); (b-3) condition wherein a plastic plate was placed between the sole and the pressure distribution sensor (Condition (C): plastic). The experimental participants were equipped with motion-capture markers and performed natural squat exercises on a pressure distribution sensor. Foot pressure distribution data and motion capture data were collected during the exercise.
}
\end{figure*}

This study was approved by the ethical committee for human studies at the Graduate School of Engineering Science, Osaka University, Japan. All experiments were performed according to the approved design. Before the experiment, all participants signed the informed consent form. The participants included in the figures provided informed consent for the publication of images in online publications.

The experimental concept is shown in Fig.\ \ref{fig:concept}. A sheet-shaped distributed pressure sensor was used to measure the pressure distribution on the plantar surface of the foot, while motion capture was used to obtain the posture and joint angles of the human body. The posture and joint angles were estimated using a multivariate linear regression model, with the pressure distribution information as the model variables. Additionally, this study investigated whether foot morphology affects the estimation accuracy. Specifically, the investigation focused on whether the estimation accuracy changes when the effect of foot morphology is eliminated by placing a rubber plate or plastic plate between the foot pressure sensor and the plantar region.

The human posture and joint angles were measured by motion capture (V120 Trio, OptiTrack, Acuity Inc.) using optical markers. The markers were attached to the participants’ ankles, knees, hips, and shoulders. The movements of the markers from the side of the participant were recorded by motion capture, and each joint angle (ankle, knee, and hip joint) and upper body posture was calculated from the positional information of the markers in the sagittal plane. Thus, this study focused only on two-dimensional motion. Specifically, the ankle joint angle, knee joint angle, and hip joint angle were calculated from the marker positions, and the upper body angle was calculated from these angles.

To measure the plantar pressure distribution, an electromagnetic induction-based pressure distribution sheet sensor (LL Sensor, LL480$\times$480, Newcom Inc.) was used to measure the participants’ foot pressure distribution. This pressure distribution sensor consists of 48$\times$48 pixels, with each pixel measuring 10$\times$10 mm. Comprehensive pressure data were collected for both foot regions.

Seven adult males participated in the experiment. The average height of the participants was 170 cm, and the average weight was 65 kg. Motion-capture markers were placed on the participants, who were subsequently instructed to perform natural squat exercises on a pressure distribution sensor with a period of 2 s. A metronome was used to adjust the pace of exercises. Measurements were taken over 30 s. The participants were instructed to refrain from changing their plantar position during measurement, to keep the entire plantar surface of their foot on the ground, and to refrain from lifting the sole off the foot pressure distribution sensor. During the exercise, foot pressure distribution data and motion capture data were collected.

The experiment was conducted under three conditions. Under the first condition, nothing was placed between the sole and the pressure distribution sensor (Condition (A): nothing). Under the second condition, a rubber plate was placed between the sole and the pressure distribution sensor (Condition (B): rubber). Under the third condition, a plastic plate was placed between the sole and the pressure distribution sensor (Condition (C): plastic). Condition (B) was set to test the effect of uniform foot stiffness characteristics on the estimation accuracy. The rubber plate was made of silicone rubber (Ecoflex0030, Smooth-On) with characteristics resembling those of the human skin and a thickness of 3 mm. Condition (C) was set to test whether the estimation accuracy is affected when the foot structure resembles a link mechanism. The 3-mm thick plastic plate was fabricated using a 3D printer with poly-lactic acid (PLA) filament. Condition (A) was implemented to verify that this method can accurately estimate joint angles and postures, while conditions (B) and (C) were implemented for comparison with condition (A). Each participant was tested five times for condition (A), three times for condition (B), and three times for condition (C), amounting to a total of 11 experiments. Because there were seven participants in the experiment, 77 datasets were obtained in total.

In the processing of experimental data, the sampling period of the pressure distribution sensor was not constant, and the data varied. Therefore, data resampling was carried out at regular intervals (20 ms) using linear interpolation. Here, 20 ms was the average sampling time of the pressure distribution sensor. Similarly, the motion capture data obtained at the sampling rate of 120 Hz were resampled to the sampling time of 20 ms. The measurement data were split into training and validation data. Specifically, the first 3 s of the measurement data were removed, and the remaining 27 s were used as training and validation data at the ratio of 5:1.

The multivariate linear regression model was used to estimate the joint angle and posture from the pressure distribution information obtained in this study. Here, the multivariate linear regression model was used because of its high interpretability, simplicity, and low learning cost compared with other machine learning models, such as neural networks. In the process of learning the weights for linear regression, the models were calculated using Ridge regression to avoid overlearning and multicollinearity problems. Lasso regression and elastic net regression were performed on some data, but the accuracy was similar to or lower than that of Ridge regression. Unlike Ridge regression, which can be optimized with analytical expressions, Lasso and elastic net regression require iterative calculations for optimization. In this study, Ridge regression was selected owing to its advantages of simplicity and speed. The specific procedure is described below.

First, variable selection was performed in the training process, because using all pressure data would include unnecessary variables, and the estimation accuracy would decrease because of overtraining. Only pressure data with high correlation coefficients and joint angles were used as model variables. Specifically, as a univariate feature selection for the Filter method, features with low correlation coefficients were excluded. The correlations with the four angles (ankle joint, knee joint, hip joint, and upper body) were calculated, and only variables with a correlation coefficient R$>$0.15 were selected. Consequently, the same set of variables was used to estimate each of the four angles. The threshold value of 0.15 was selected by grid search. This study selected the value at which the average value of R2 for all experiments under condition (A) was highest. The results reveal that there was almost no difference in the mean value of R2 at the threshold values in the range 0.10$<$ R$<$0.25. Subsequently, time series data for the pressure sensors (all pressure data selected by variable selection), and angles (all joints and posture angles) were standardized (Z-score normalization) for training. Thus, the selected and standardized pressure data $p^i_k$ at the k-th time data point and i-th location and each standardized joint data $\theta_k$ at the k-th time data point were collected. 

Subsequently, the multivariate linear regression model was used to estimate the joint angle and posture from the pressure distribution information, as follows. Notably, the following description pertains to a method for estimating a single joint angle or posture in a single experiment for one experimental dataset.

Equation (\ref{eq:1}) was used to calculate the estimated joint angle at the k-th time data point in the multivariate linear regression model. 

\begin{eqnarray}
\hat{\theta}_k = \sum_{i} w^i  p^i_k
\label{eq:1}
\end{eqnarray}

where \textit{$\hat{\theta}_{k}$} is the estimated joint angle at the k-th time data point, \textit{$p_k^{i}$} represents the pressure data of the i-th location at the k-th time data point, and \textit{$w^i$} is the weight corresponding to the pressure data at the i-th location.

A linear regression model was trained using Ridge regression, which is a method for calculating the weight \textit{$\mathbf{w}=(w^{1},w^{2}, \cdots)$} that minimizes the following loss function with the residual sum of squares term and the L2 regularization term. 
\begin{eqnarray}
L=\sum_{k}\left(\theta_k-\hat{\theta}_k\right)^2+\frac{1}{2} \lambda \sum_{i} (w^{i})^2
\label{eq:a1}
\end{eqnarray}
where \textit{$\theta_{k}$} is the measured joint angle at each time data point k.

The weight \textit{$\mathbf{w}$} minimizing the loss function was determined analytically using the following equation.

\begin{eqnarray}
\mathbf{w}= \boldsymbol \theta  \mathbf{P}^{\mathrm{T}} \left( \mathbf{P} \mathbf{P}^{\mathrm{T}}+\lambda \mathbf{I}\right)^{-1}
\label{eq:a2}
\end{eqnarray}
where 
\textit{$\mathbf{w}=(w^{1},w^{2}, \cdots)$} is the weight vector; 
\textit{$\boldsymbol \theta = \left (\theta_{1}, \theta_{2}, \cdots   \right )$} is the joint vector summarizing the measured joint angle at each time data point;\textit{$\mathbf{P}= (\mathbf{p^{1}},\mathbf{p^{2}}, \cdots)$} is the pressure state matrix summarizing the vectors of the measured pressure data; \textit{$ \mathbf{p^i} = \left (p^{i}_1,p^{i}_2, \cdots   \right )^T $} is the pressure vector of the i-th location;
\textit{$\lambda $} is a hyperparameter and \textit{$\mathbf{I}$} is the identity matrix. The hyperparameter \textit{$\lambda $} was determined by trial and error, and the same value (\textit{$\lambda=10$}) was used for all training.

Using the above-mentioned methods, estimations were obtained at all joint angles and postures for all trials. Notably, the learning and estimation processes were carried out for each joint angle and posture in each dataset of each experiment, and a different learner was used in each case.

For each dataset, the measured and estimated joint angles were subjected to the inverse transformation of standardization, and their original scale was restored. Then, the Root Mean Square Error (RMSE) and coefficient of determination (R2) values between the estimated and measured values in the validation data were calculated. 

Additionally, statistical analyses were performed on the RSME and R2 results between the data obtained under experimental conditions (A) and (B), and between the data obtained under experimental conditions (A) and (C). In these statistical analyses, the Shapiro-Wilk test revealed normality in the data, and Welch’s t-test was performed because the F-test did not reveal equal variance in the data. The statistical significance was set to p$<$0.05. The reason for performing statistical analysis between the data for experimental conditions (A) and (B) was to test whether the intervention of the rubber plate, which ensures uniform sole characteristics, has a significant influence on the estimation accuracy. Statistical analysis was also conducted for the data obtained under experimental conditions (A) and (C) to test whether there is a difference in the estimation accuracy when an intervening plastic plate makes the foot structure resemble a link structure.

\section{\label{result} Results}

\begin{figure*}
\includegraphics[width=14cm]{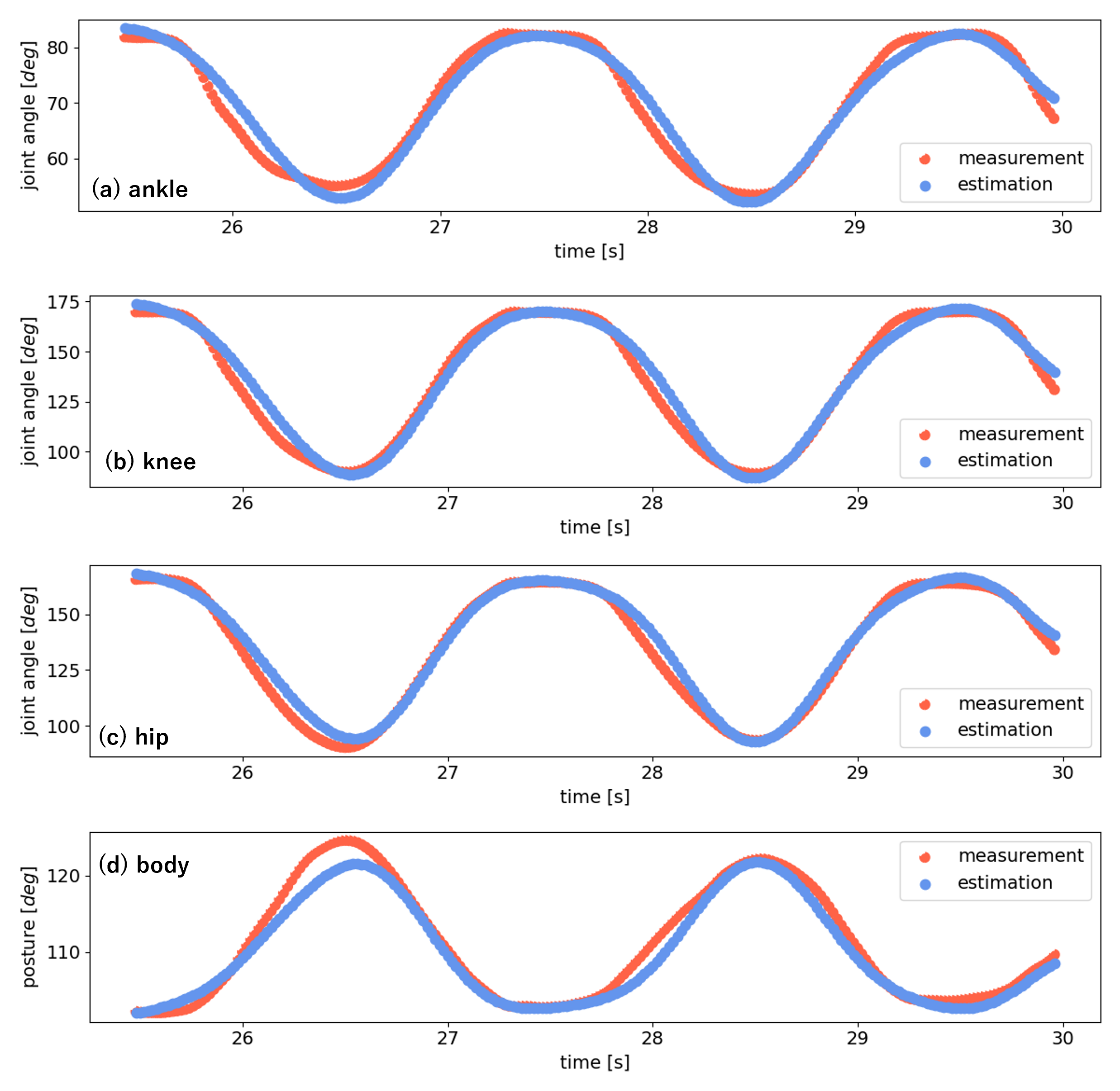}
\caption{\label{fig:time} 
Representative example of estimation results for: \textbf{(a)} ankle joint angle, \textbf{(b)} knee joint angle, \textbf{(c)} hip joint angle, and \textbf{(d)} upper body posture under experimental condition (A). This figure shows only the range of validation data. For comparison, the plots for each output (a)–(d) are overlaid with the time series of the measurement data \textit{$\theta$} (red line) and estimated data \textit{$\hat{\theta}$} (blue line). These results reveal the high estimation accuracy for the joint angles and posture using the foot pressure distribution.
} 
\end{figure*}

\begin{figure*}
\includegraphics[width=14cm]{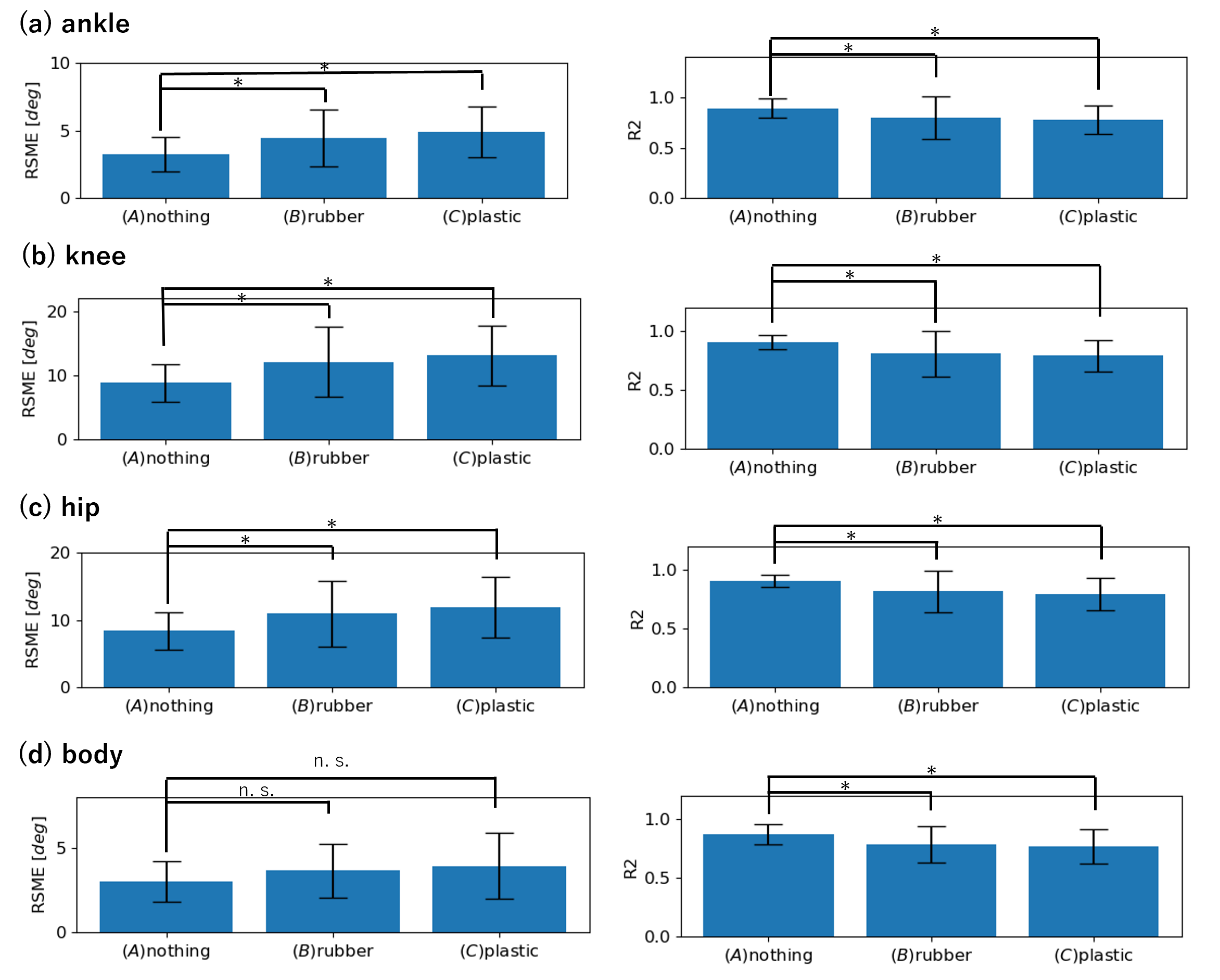}
\caption{\label{fig:stats} 
Results for error (RMSE and R2) in estimation of \textbf{(a)} ankle joint angle, \textbf{(b)} knee joint angle, \textbf{(c)} hip joint angle, and \textbf{(d)} upper body posture under each experimental condition. The figure also shows the statistical analyses performed on the RSME and R2 results for the data obtained under experimental Condition (A) and (B), and the data obtained under experimental Conditions (A) and (C). The statistical significance was set to p$<$0.05. The difference in the measurement accuracy between cases with and without an intervening object is significant, with lower estimation accuracy for the rubber plate (Condition (B)) and plastic plate (Condition (C)) compared with the case without an intervening object (Condition (A)), which is true for all results except the RSME of body posture estimation. 
} 
\end{figure*}

Figure \ref{fig:time} shows a representative example of the estimation results for the joint angles of the ankle, knee, and hip joints, and the upper body posture under experimental condition (A). Figure \ref{fig:stats} shows the estimation error (RMSE and R2) results for each experimental condition. The results for experimental condition (A) in Fig. \ref{fig:stats} reveal that the accuracy of estimating the angle of each joint and the posture of the upper body using the foot pressure distribution is R2$\fallingdotseq$0.9. The RMSE was estimated at 3.3 $\pm$ 1.3 deg, 8.8 $\pm$ 3.0 deg, 8.4 $\pm$ 2.8 deg, and 3.0 $\pm$ 1.2 deg for the ankle joint, knee joint, hip joint, and upper body posture, respectively.

Figure \ref{fig:stats} summarizes the estimation results for conditions (A), (B), and (C), respectively. The figure shows the mean and variance of the estimation error (RMSE and R2) for each experimental condition. From Figure \ref{fig:stats}, the results for Condition (A) have a lower RMSE compared with the other results. Additionally, the results for Condition (A) have a higher R2 compared with the other results.

The results obtained by the statistical analysis of the ankle joint, knee joint, hip joint, and upper body posture are shown in Fig. \ref{fig:stats}. The results obtained by the statistical analysis of the estimation results for the ankle joint, knee joint, and hip joint are discussed below. For all estimations, there are significant differences in the RMSE and R2 between condition (A) and condition (B). When the rubber plate was interposed at the foot, the estimation accuracy was significantly lower. Similarly, there are significant differences between condition (A) and condition (C) in all estimations. Therefore, the estimation accuracy was significantly reduced when the plastic plate was interposed at the foot. The results obtained by the statistical analysis of the upper body posture are discussed below. The difference between condition (A) and condition (B) is significant for R2 (p=0.03), but not for RMSE (p=0.12). The difference between condition (A) and condition (C) is significant for R2 (p=0.008), but not for RSME (p=0.07).

\section{\label{discussion} Discussion}

A highly accurate estimation of R2$\fallingdotseq$0.9 was achieved in the estimation of the joint angles and posture using foot pressure distribution information. These results reveal that the posture and joint angles of the human body can be estimated with high accuracy from foot pressure distribution information using multivariate linear regression. The estimation accuracy achieved by a previous study that used image-based estimation for squatting motion similar to that in this study are as follows: upper body posture error: 2.8 $\pm$ 0.6 deg; knee joint angle error: 8.0 $\pm$ 1.4 deg \cite{lin2022automatic}. The estimation results obtained by the proposed method are comparable. Therefore, the proposed method can estimate joint angles and posture with sufficient accuracy for some applications.

To estimate multiple postures and joint angles, the proposed method only requires that a pressure distribution sensor is attached to the plantar part of the foot, which is the end of the body in contact with the environment. Additionally, the proposed method is characterized by its ability to estimate multiple angular information of the human body from a single source, that is, foot pressure distribution information, and can also obtain information for locations away from the foot, such as the angle of the knee joint, hip joint, and upper body. Additionally, sensors are only required for the foot sole, and there is no need to attach sensors near the joints of the human body. Generally, when using contact and wearable sensors, multiple sensors must be placed near each part of the body to estimate multiple joint angles and postures. A distinctive feature absent in other contact and wearable sensing methods is the ability to estimate joint angles at a distance from the sensor, eliminating the need to place sensors outside of the foot.

The statistical analysis results indicate that the plantar morphology contributes to the joint angle and posture estimation. There is a difference in the measurement accuracy between cases with and without an intervening object. Specifically, the estimation accuracy is lower for the rubber plate and plastic plate. The rubber plate is considered to have the effect of equalizing the hardness distribution in the plantar region, while the plastic plate is considered to make the structure of the plantar region inconsequential, giving it a morphology resembling a link structure. These results suggest that the morphology (structure and characteristics) information of the foot is required to perform the proposed estimation. In the field of biological systems, it is important to understand how the human body functions in terms of perception and movement. The plantar region is exposed to the environment and receives various information from it \cite{meyer2004role}. The plantar region perceives motion and plays a role in motor control \cite{okoba2019particular}, and plantar skin information is known to play an important role in human postural control \cite{kavounoudias1998plantar}. The morphology of the human foot is very complex \cite{lai2022evaluating}, and the extent to which the foot sensory system is involved in motor control remains unclear \cite{meyer2004role}. The results of this study are interesting from the perspective of morphological computation. In this field, the morphology of the human body is considered to have certain computational functions \cite{Pfeifer2007how}. The results of this study reveal that the joint angle and posture estimation from foot pressure distribution information ceases to function when the ground contact conditions change (Condition (B) and (C)). Therefore, the morphology of the plantar region functions as a type of computational function and contributes to the estimation of the joint angle and posture. Here, the complex morphology of the foot plays an important role in the generation of the complex and diverse information of the foot pressure distribution, and this complex and diverse information is required for estimation. Moreover, the results indicate that it is possible for the complex morphology of the human foot to function as morphological computation and serve as a computational resource for human perception and motor control. Notably, however, these conclusions do not have a physiological and anatomical basis, and are only based on empirical and indirect data.

In recent years, studies on physical reservoir computing for a soft body, which considers the dynamics of soft materials as information processing devices, such as recurrent neural networks, have reported the possibility of using various physical phenomena as computational resources \cite{tanaka2019recent, nakajima2020physical}. Some studies have investigated the computational capabilities of the musculoskeletal system \cite{sumioka2011computation}, used the soft spine as a physical reservoir to generate the motion of a quadruped robot \cite{zhao2012embodiment}, processed information using a soft body such as an octopus \cite{nakajima2015information, nakajima2014exploiting, nakajima2013soft, nakajima2018exploiting}, and used soft material bodies as computational resources. Another study proposed a soft wearable suit with tactile sensors, which can be considered as a tactile sensor network for monitoring natural body dynamics, as a computational resource for estimating human and robot posture \cite{sumioka2021wearable}. In physical reservoir computing, the intermediate layer uses nonlinear physical states, and the output layer is a linear and static readout. Then, only the output layer is learned. The output layer and learning process used in physical reservoir computing are the same as those in the multivariate linear regression model and its learning method used in this study. Therefore, the method proposed herein can be considered as an information processor that introduces the concept of physical reservoir computing and uses the human foot as a computational resource to estimate the joints and posture. In other words, the foot dynamics are considered as a physically implemented recurrent neural network, and the joint angles and posture are estimated from the time series data of the foot state (pressure sensor data) using only linear and static readouts (multivariate linear regression model). This concept, which uses the human foot as a computational resource, contributes toward highly accurate estimation with a simple model and learning method, and toward the significant reduction of training and computation cost. We name these types of research area ``biomechanical reservoir computing," which utilizes the physical dynamics occurring in the human (organisms) interior and in contact with surrounding environment as a computational resource, and are promoting researches such as \cite{kobayashi2023information} using this framework.

\begin{figure*}
\includegraphics[width=14cm]{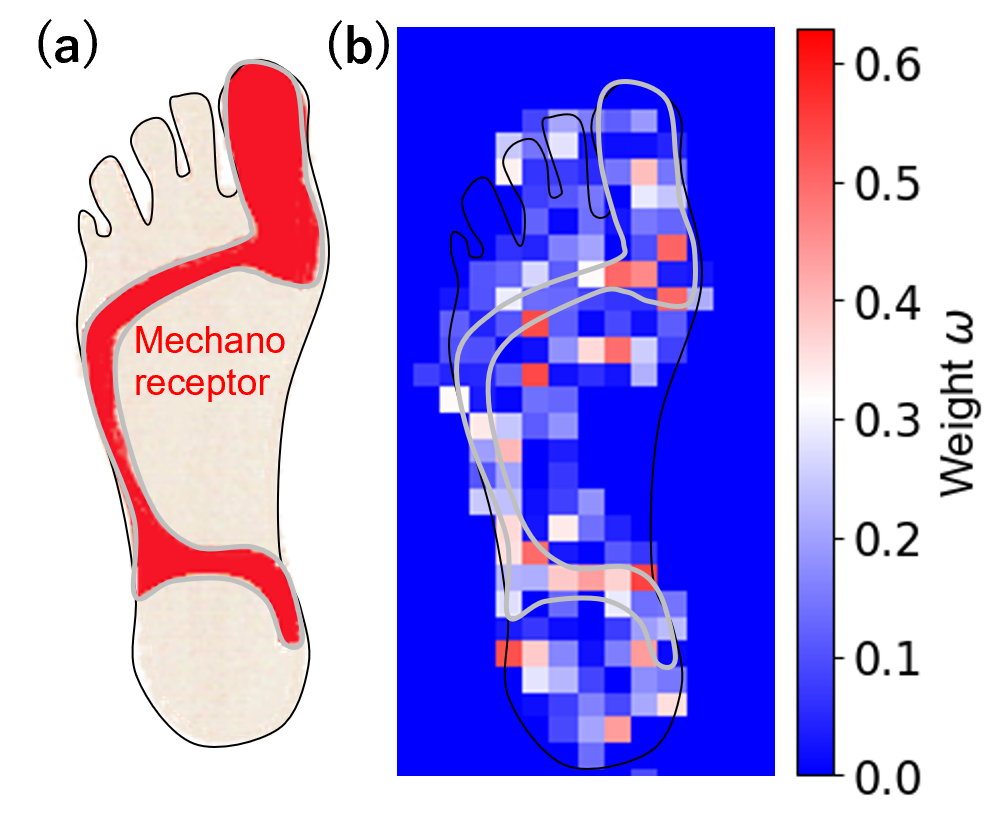}
\caption{\label{fig:map} 
\textbf{(a)} Locations where mechanoreceptors are considered to be abundant in the plantar region and \textbf{(b)} representative example of map of absolute value of weight $\omega$ in multivariate linear regression model for each pressure sensor location. By comparing (a) and (b), the locations where the information weights in the estimation method are large, and the locations where there are many mechanoreceptors in the plantar region, are qualitatively consistent to some extent. 
} 
\end{figure*}

Consideration should be given to the information processing methods of this study in terms of their relationship with mechanoreceptors. For the information processing in this study, a very simple model was used as the multivariate linear regression model, in which the time series data for each pressure are weighted and added together. Notably, because all pressure data are standardized, the pressure magnitude at a given location does not affect the estimation accuracy, which is only affected by the waveform pattern of the time series data. Figure \ref{fig:map}(b) shows a representative example of a map of the absolute value of weight $\omega$ in the multivariate linear regression model for each pressure sensor location. Here, when the absolute value of $\omega$ in the estimation method is large, the information at the pressure sensor location makes a large contribution to the estimation. Figure \ref{fig:map} shows that the weight $\omega$ is large at locations where mechanoreceptors are considered to be abundant in the plantar region, such as the thumb and base of the five toes \cite{Sensomotorische2000Bizzini,Kirsten2006Gehen}. Figure \ref{fig:map}(a) shows information for locations where mechanoreceptors are considered to be abundant in the plantar region \cite{Sensomotorische2000Bizzini,Kirsten2006Gehen}. By comparing Figure \ref{fig:map}(a) and Figure \ref{fig:map}(b), it can be seen that the locations where the weights of the information in the estimation method are large, and the locations where there are many mechanoreceptors in the plantar region, are qualitatively consistent to some extent. This suggests that, in the perceptual processing of organisms, information may be processed by a mechanism resembling that of the proposed estimation method. In other words, it is hypothesized that, in the perceptual processing of living organisms, the joint angles and posture are estimated as proprioceptive sensations by integrating the local time-series patterns of pressure at locations where there are many mechanoreceptors in the plantar region. Notably, however, these conclusions are not based on quantitative investigation or physiological data, and there is only qualitative agreement regarding the location.

Hence, the proposed framework may also help in elucidating the role of the human body and other living organisms. Morphological computation and physical reservoir computing with a soft body can be considered as an attempt to harness the inherent computational capabilities of living organisms. Investigating which computational tasks can be performed using living organisms will provide new perspectives for understanding questions such as why living organisms have such specialized morphology. Whether living bodies can serve as morphological computation or physical reservoir computing, and whether they can serve as computational resources for biological perception and motor control has been a matter of debate \cite{kobayashi2023information, nonaka2020locating}. The findings obtained by this study can provide interesting research directions for further investigation.

The proposed method for estimating posture and joint angles using foot pressure distribution is expected to find application in the estimation of the angle of each joint during walking. By using only one plantar pressure distribution sensor installed in the shoe, it may be possible to estimate the angles of multiple joints during walking, eliminating the need for additional sensors. However, as the foot lifts off the ground during the stance and the ground contact conditions between the plantar area and the pressure distribution sensor change, the learner may need to change depending on the gait phase.

The experimental results in Fig.\ref{fig:time} reveal that the waveform patterns of the ankle, knee, and hip joints are similar during squatting movement. Therefore, it is possible that there is an excess of joint types that can be estimated. Future work should investigate whether similar results can be obtained for non-natural movement with different waveform patterns.

Notably, because the surface material of the pressure distribution sensor used in this study was flexible, even when a plastic plate was interposed between the foot and the sensor, a shift in the pressure distribution caused by plastic plate tilt was observed. This flexibility of the pressure distribution sensor may be a factor affecting the measurement accuracy when a plastic plate exists, and this may have influenced the results of the statistical analysis of the upper body posture, which did not reveal significant differences in the RMSE between condition (A) and condition (B), and between condition (A) and condition (C).

\section*{data availability}

The datasets generated in this study are available from the corresponding author upon reasonable request. The data are not publicly available to protect the privacy of the research participants.

\begin{acknowledgments}
This study was supported in part by a Grant-in-Aid for Scientific Research from the Ministry of Education, Culture, Sports, Science and Technology (MEXT) (19H02112, 19K22878, and 23K18479), Japan. This study was supported in part by JST, PRESTO (Grant Number JPMJPR23P2), Japan. This study was supported in part by Innovation Inspired by Nature, Research Support Program, SEKISUI CHEMICAL CO., LTD. 
\end{acknowledgments}


\bibliographystyle{unsrt}
\bibliography{Manuscript} 

\end{document}